\documentclass[conference]{IEEEtran}

\usepackage{cite}
\usepackage{amsmath,amssymb,amsfonts}
\usepackage{graphicx}
\usepackage{booktabs}
\usepackage{multirow}
\usepackage{xcolor}
\usepackage{url}
\usepackage{tikz}
\usetikzlibrary{arrows.meta, positioning, fit, calc}

\graphicspath{{../paper/figures/}{./figures/}{./}}

\newcommand{\encoder}[1]{\textsc{#1}}

\begin{document}

\title{Taste-aware music retrieval from audio embeddings}

\author{%
  \IEEEauthorblockN{Matteo Spanio}
  \IEEEauthorblockA{%
    \textit{University of Padua}, Padua, Italy\\
    spanio@dei.unipd.it
  }%
  \and
  \IEEEauthorblockN{Antonio Rodà}
  \IEEEauthorblockA{%
    \textit{University of Padua}, Padua, Italy\\
    roda@dei.unipd.it
  }
}

\maketitle

\begin{abstract}
Crossmodal correspondences between sound and taste are well established in psychology and neuroscience, but largely absent from content-based multimedia retrieval. We formalise taste-from-audio prediction as a content-based music information retrieval benchmark over a perceptually validated multi-source corpus, comparing ten frozen audio encoders from the four HEAR families under a shared multi-task regression head, with gated late-fusion as a configurable variant. In order to assess the effectiveness of the models, we compute absolute error and rank correlation. The strongest systems predict the five tastes within a macro RMSE of $0.134$; on held-out real music their error is less than half a single rater's deviation from the consensus (RMSE $0.13$ vs.\ $0.28$), so the model tracks the group consensus more closely than an average human rater, and well below the previous state of the art baseline ($0.219$). On absolute error the encoders are statistically flat, with a single \encoder{VGGish} matching the best fusion, but gated late-fusion's advantage is confined to rank correlation (macro Pearson $r$ $0.724$ vs.\ $0.666$). Operationalised as a content-based retrieval index, the predicted taste space ranks a $309$-item pool far more faithfully than a CLAP-text baseline, which sits at chance; ridge probes and an audio-bandstop knockout read the strongest representations against documented sound--taste correspondences.
\end{abstract}


\section{Introduction}

Crossmodal correspondences between sound and taste are one of the clearest examples of stable associations between audition and the chemical senses. High pitch, consonance, and bright timbre are repeatedly associated with sweetness, while lower pitch, roughness, and darker timbre shift listeners toward bitterness and sourness \cite{spence2011crossmodal,knoferle2012sounds,wang2017assessing}. These effects already motivate sonic-seasoning applications in restaurants, advertising, and multisensory design \cite{spence2017gastrophysics}, yet they remain peripheral to mainstream music information retrieval (MIR) and multimedia benchmarking.

This gap matters for content-based multimedia indexing. If taste judgments can be predicted from audio in a reproducible way, they become a new semantic axis for organising collections, querying music beyond genre and mood, and recommendation scenarios such as ``a similar track but sweeter''. The task also creates an unusual test-bed for explainable multimedia learning: a model is useful only if it scores well \emph{and} if its behaviour can be compared with empirical findings from psychology and neuroscience.

\begin{figure}[!t]
\centering
\includegraphics[width=\columnwidth]{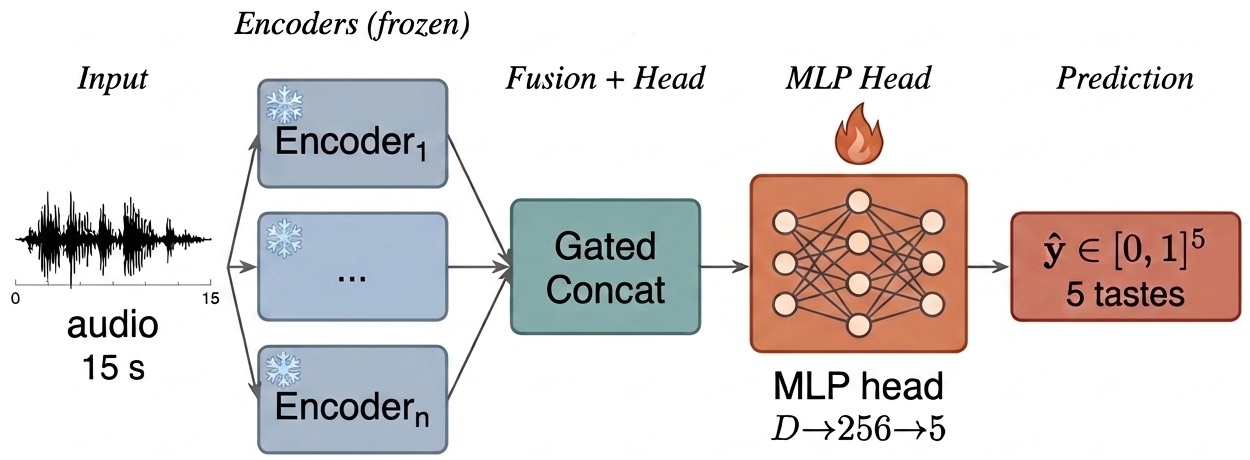}
\caption{Proposed model architecture. One or more frozen audio encoders produce per-encoder embeddings that are concatenated and re-weighted by a learned per-encoder gate; the gated representation feeds a shared two-layer MLP whose sigmoid head outputs a 5-D taste vector in $[0,1]^5$.}
\label{fig:architecture}
\end{figure}

Taste tagging from audio is not a completely novel task. Guedes et al.\ introduced the Taste \& Affect Music Database \cite{guedes2023taste}; Rodriguez fine-tuned five separate Audio Spectrogram Transformer (AST) regressors, one per taste, on a curated $257$-song soundtracks corpus and used them to label the FMA corpus at scale \cite{rodriguez2024sonic}; Spanio et al.\ extended the corpus, validated the labels perceptually, and explored generative variants \cite{spanio2025multimodal,spanio2026multimodal}. Prior work established the feasibility of taste prediction from audio, with evaluation centred on direct regression error. The present paper expands it into a broader content-based MIR setting; we make three contributions.
\begin{itemize}
    \item We formalise taste-from-audio as a content-based MIR benchmark over a perceptually validated multi-source corpus, with a frozen-encoder protocol across ten encoders covering the four HEAR families~\cite{turian2022hear}, in single-encoder and gated late-fusion forms (five-seed means). Read for rating replacement, the best systems reach a macro RMSE of $0.134$; on held-out real music this is less than half a single rater's error to the consensus (RMSE $0.13$ vs.\ $0.28$) and far below the previous baseline~\cite{rodriguez2024sonic} ($0.219$). A single \encoder{VGGish} already reaches this error; gated late-fusion adds only rank correlation ($r$ $0.724$ vs.\ $0.666$).
    \item We operationalise the predicted taste space as a content-based retrieval index: over a $309$-item pool with taste-profile queries~\cite{spanio2026multimodal} it leads every metric, while a generic CLAP-text~\cite{wu2023clap} baseline collapses to chance, to our knowledge the first retrieval evaluation of taste-conditioned music indexing.
    \item We pair the benchmark with a psychophysics-grounded interpretability layer, ridge probes to nine spectral descriptors and a cross-encoder audio-bandstop knockout, reading the strongest representations against documented sound--taste correspondences~\cite{spence2011crossmodal,knoferle2012sounds,wang2017assessing}.
\end{itemize}

\section{Related Works}

\subsection{Sound-taste correspondences and sonic seasoning}

Crossmodal correspondences between auditory parameters and basic tastes have been documented for over a decade. Psychophysical work maps pitch, consonance, timbre, and tempo onto the four canonical tastes: high pitch, consonance, and bright timbre with sweetness, low pitch, roughness, and darker timbres with bitterness \cite{spence2011crossmodal,knoferle2012sounds,crisinel2010crossmodal,wang2017assessing}, sweet and bitter showing the largest effects, with replications confirming the instrument-to-taste mapping \cite{wang2016multisensory,wang2021multisensory,qi2020perception,watson2017multisensory}. Sound also modulates perceived intensity, texture, and emotional appraisal of food and drink \cite{zampini2010sound,mathiesen2022sound,galmarini2021impact}, grounding the sonic-seasoning program in restaurants, advertising, and multisensory design \cite{spence2017gastrophysics}. These are \emph{crossmodal correspondences}, population-wide sound--taste associations distinct from synaesthesia's idiosyncratic cross-activation in a small minority \cite{spence2011crossmodal}; we model the former, shared across listeners.

Computational work on sound--taste has developed along four threads. \emph{Compositional}: \cite{mesz2012composition} introduced a music-generation algorithm built on taste--music correspondences. \emph{Stimulus norming}: \cite{guedes2023taste} produced a 100-track music database normed for sweet/bitter/sour/salty. \emph{Discriminative}: \cite{rodriguez2024sonic} fine-tuned five separate Audio Spectrogram Transformer regressors on a curated soundtracks corpus and used them to label the FMA dataset \cite{defferrard2017fma} at scale. \emph{Generative and dataset-level}: Spanio et al.\ surveyed the multimodal-generative-AI design space for sound and taste \cite{spanio2024towards}, fine-tuned MusicGEN \cite{copet2024simple} on a taste-prompted corpus \cite{spanio2025multimodal}, and produced a unified music--taste corpus validated by a 49-participant perceptual study \cite{spanio2026multimodal}. The present paper sits in the discriminative thread, replacing the five-head per-taste architecture with a single multi-task head over frozen encoders.

\subsection{Audio representations, explainability, and retrieval}

Modern audio representations span supervised event models~\cite{hershey2017vggish,kong2020panns,gong2021ast}, self-supervised speech and music models ~\cite{hsu2021hubert,li2024mert,alonsojimenez2024omarrq}, multimodal and codec-based models~\cite{wu2023clap,defossez2023encodec}, and music-pretrained representations ~\cite{mccallum2022mule}. Benchmarks such as HEAR \cite{turian2022hear} aggregate tasks across these families but exclude sensory dimensions like taste; \cite{kang2024music} reaches a similar conclusion for music emotion. Closest to us, \cite{frommholz2026predicting} predicts 19 perceived-semantic axes of functional sounds by feature extraction and ensembles, and \cite{lepa2020computational} predicts perceived musical expression on advertising stimuli, both from audio features rather than human-rated taste labels.

For explainability, ridge probes over embeddings~\cite{frommholz2026predicting} and Centered Kernel Alignment (CKA)~\cite{kornblith2019cka} are standard probes of what an embedding linearly preserves; we adopt both. For attribute-guided retrieval,~\cite{wilkins2026controllable} showed that frozen music embeddings support attribute-controlled search through a learned transformation; we treat their setup as the downstream test bed and report a static retrieval-by-taste-profile baseline robust at our corpus scale. 
\section{Methodology}

\subsection{Dataset and task}

Experiments use the unified music--taste corpus of~\cite{spanio2026multimodal}, which aggregates audio--taste annotations from three sources, through a fixed split column that we do not reshuffle (269 train + 68 val + 40 test clips): the real-music perceptually validated subset of~\cite{spanio2026multimodal} (49 raters, 20 dishes), the MusicGen-generated subset of~\cite{spanio2025multimodal} (436 raters, 100 songs), and the soundtracks survey corpus of~\cite{rodriguez2024sonic} (257 songs). Targets are five normalised taste intensities (sweet, bitter, salty, sour, spicy): the first four are the canonical basic tastes whose crossmodal mappings to sound are characterised in psychophysics~\cite{spence2011crossmodal,knoferle2012sounds,wang2017assessing,guedes2023crossmodal}, while spicy is a trigeminal/chemesthetic axis rated alongside them in the source studies, though less constrained by prior evidence. Each clip carries partial observation masks (the MusicGen-generated subset omits spicy), so we train with a masked MSE loss to avoid phantom-zero signals on unobserved cells. Because subjective labels carry corpus-specific noise~\cite{kang2024music,frommholz2026predicting}, we report metrics ceiling-aware: the test set mixes 20 real-music~\cite{spanio2026multimodal} and 20 MusicGen-generated items~\cite{spanio2025multimodal}, whose leave-one-out inter-rater Pearson ceilings differ substantially ($\approx 0.60$ vs.\ $0.25$), so each model is read against the noise floor of its source.

\subsection{Models architecture}

The model is a frozen audio encoder followed by a small multi-task head (Fig.~\ref{fig:architecture}). For any 15\,s audio clip $\mathbf{x}$, encoder $f_\theta$ produces an embedding $f_\theta(\mathbf{x}) \in \mathbb{R}^D$ that is mean-pooled along the time axis and cached on disk; a two-layer MLP (hidden size 256, ReLU, dropout $0.2$) with a final sigmoid maps the embedding to a 5-D taste vector in $[0,1]^5$. Frozen-encoder + linear/MLP probes are the standard recipe for benchmarking pretrained audio representations on small downstream corpora~\cite{turian2022hear,frommholz2026predicting}; we adopt it for reproducibility and because cached embeddings keep an encoder sweep tractable at this scale. The sigmoid bounds the output and prevents the out-of-range predictions that the previous SOTA's five unbounded linear regressors produce on the same data~\cite{rodriguez2024sonic}. Masked MSE keeps the partial labels of the MusicGen subset from collapsing to phantom zeros, and the multi-task head exploits the strong inter-target correlations in normed taste--music corpora~\cite{guedes2023taste,spanio2026multimodal}.

\textbf{Training.} AdamW, learning rate $10^{-3}$, weight decay $10^{-4}$, batch size $32$, up to $50$ epochs with early stopping (patience $10$) on the validation macro $r$. Every reported metric for the multi-task MLP head and its fusion variants is the mean over five seeds ($\{11, 22, 33, 44, 55\}$); the encoder cache is computed once and reused across seeds.

\subsection{Audio encoders and fusion}

The benchmark spans ten frozen audio encoders that together cover the four families surveyed in~\cite{turian2022hear}: supervised event/scene models (VGGish~\cite{hershey2017vggish}, PANNs~\cite{kong2020panns}, AST~\cite{gong2021ast}), self-supervised speech and music models (HuBERT~\cite{hsu2021hubert}, MERT~\cite{li2024mert}, Omar-RQ~\cite{alonsojimenez2024omarrq}), multimodal and codec-based models (CLAP~\cite{wu2023clap}, EnCodec~\cite{defossez2023encodec}), and music-pretrained representations (MULE~\cite{mccallum2022mule}), with MFCC as the DSP floor. Frozen encoders are the default regime because the task is low-$N$ and cached embeddings make broad comparison reproducible. We also evaluate the previous SOTA \texttt{ast-5head-ref}~\cite{rodriguez2024sonic}, five fine-tuned AST regressors (one per taste) on raw audio, the natural reference for our single multi-task head.

\textbf{Gated late-fusion.} To combine encoders we add a learned gate over the concatenated embeddings: each per-encoder slice is scaled by a sigmoid weight before the MLP head, the gate trained jointly with the head while encoders stay frozen. We pick fusion candidates by complementarity rather than raw rank, following the ensemble logic of~\cite{frommholz2026predicting} and the cross-objective combination of~\cite{wilkins2026controllable}, and evaluate seven configurations over the four strongest encoders: the pairs AST$+$VGGish, AST$+$MULE, VGGish$+$MULE, CLAP$+$MULE, CLAP$+$VGGish, the triple AST$+$VGGish$+$MULE, and the four-way CLAP$+$AST$+$VGGish$+$MULE.

\subsection{Explainability and retrieval probes}

We attach two analysis layers to the strongest models. First, \emph{psychoacoustic ridge probes}: a 5-fold cross-validated regression from each cached embedding to nine acoustic descriptors (spectral centroid, rolloff, bandwidth, flatness, ZCR, RMS, tempo, harmonic-to-noise ratio, contrast), the cues routinely linked to crossmodal taste percepts (bright/sharp spectra to sweet, low-frequency energy to bitter)~\cite{spence2011crossmodal,knoferle2012sounds,crisinel2010crossmodal,watson2017multisensory,wang2021multisensory}; an embedding that still linearly preserves them has not discarded what the literature predicts a taste-aware model should use. Descriptors come from \texttt{librosa}~\cite{mcfee2015} (default $2048$/$512$ window/hop); features and targets are standardized on the train fold, ridge $\alpha=1.0$, and held-out $R^2$ reports what each embedding preserves~\cite{frommholz2026predicting}. Second, a cross-encoder \emph{audio-bandstop knockout}: a zero-phase $4$th-order Butterworth bandstop (forward--backward) at each of eight mel bands over $0$--$8$\,kHz, RMS-renormalised so loudness is not a confound, then re-encoded; the per-(band, taste) Pearson drop $\Delta r$ localises which bands the encoder relies on, in the spirit of CKA~\cite{kornblith2019cka} and frequency-band ablation~\cite{frommholz2026predicting} but applied at the input.

For the downstream task we index test items by their predicted 5-D taste vectors and rank them against five named taste-profile queries, the same regime as attribute-conditioned music retrieval~\cite{wilkins2026controllable} with taste replacing mood. The queries are not handcrafted: each is the mean perceptually rated taste vector of foods with that dominant taste in the dataset of~\cite{spanio2026multimodal} (sweet for \texttt{dessert}, bitter--sweet for \texttt{dark-chocolate}, salty for \texttt{umami-savory}, sour for \texttt{citrus}, spicy for \texttt{chili-burn}). Items rank by Euclidean distance; precision@$k$ binarises ground-truth distance at its median (applied uniformly to every system), complemented by a test-only Spearman $\rho$ and a test-vs-distractor AUC that uses no distractor labels.

\section{Experiments}

We evaluate three linked research questions, each anchored to one subsection of the Results.
\textbf{RQ1.} Under a unified multi-task head with masked MSE, which encoder family transfers best to taste prediction, and does late fusion across complementary families improve macro Pearson $r$ over the previous SOTA~\cite{rodriguez2024sonic}?
\textbf{RQ2.} Do the strongest systems preserve interpretable spectral cues consistent with the crossmodal psychophysics literature~\cite{spence2011crossmodal,knoferle2012sounds,wang2017assessing,guedes2023crossmodal}, and can the model's reliance on specific input frequency bands be localised per taste?
\textbf{RQ3.} Does the predicted 5-D taste space support content-based retrieval against principled taste-profile queries~\cite{wilkins2026controllable}, and does it outperform random ranking and text-based retrieval through a generic audio--text embedding~\cite{wu2023clap}?

\paragraph{Metrics.} For RQ1 we report per-target Pearson $r$, macro Pearson $r$, macro MAE, and macro RMSE on the held-out test split, with a per-source breakdown so absolute scores can be read against the source-specific inter-rater ceilings of~\cite{spanio2026multimodal,spanio2025multimodal}, a reporting style consistent with how~\cite{kang2024music} and~\cite{frommholz2026predicting} recommend handling subjective-label benchmarks. The metric we foreground follows the use case. Because the intended application is to replace or augment a human rater, the error metrics (RMSE, MAE, in rating units) are primary and are compared against a single human rater's leave-one-out error to the consensus, computed on the real-music subset from the 49-rater study~\cite{spanio2026multimodal}; macro Pearson $r$ is foundamental as well, reported for comparability with prior work~\cite{rodriguez2024sonic,guedes2023taste} and as the rank metric for the retrieval task (RQ3). Multi-seed configurations are reported as mean $\pm$ std over five seeds. For RQ2 we report held-out $R^2$ of the psychoacoustic ridge probes (5-fold CV, $\alpha=1$) and per-taste $\Delta r$ when each mel band is masked at the input. For RQ3 we use the five named queries described in \S\ref{sec:results-retrieval}, derived from the food-stimulus dataset of~\cite{spanio2026multimodal}, and report $P@k$ ($k \in \{5, 10, 20\}$), test-only Spearman $\rho$, and test-vs-distractor AUC against three reference points: (i) a random-ranking baseline, (ii) a CLAP-text retrieval baseline~\cite{wu2023clap} on natural-language paraphrases of each query, and (iii) the previous five-AST-heads SOTA~\cite{rodriguez2024sonic} ranking by its own predicted taste vectors. We do not include a ground-truth oracle in the table because, under the precision protocol used here, it trivially scores $1.0$ on every $k$ and carries no diagnostic information.

\paragraph{Significance.} Encoder rankings are confirmed by paired-bootstrap tests over the test split with $2000$ resamples; only the rankings explicitly described as ``ties'' fail to clear $p<0.05$. We retain source-aware reporting throughout because aggregate numbers alone hide label-noise effects on the MusicGen-generated subset~\cite{spanio2025multimodal}.

\begin{table*}[th!]
\caption{Held-out test-set metrics for ten frozen encoders, the previous SOTA five-AST-heads baseline~\cite{rodriguez2024sonic}, and seven gated late-fusion variants. Per-target Pearson $r$ in the five middle columns; macro $r$, MAE, RMSE in the rightmost block. Every learned configuration (single encoders and fusions) is the mean over five seeds $\{11,22,33,44,55\}$; \texttt{ast-5head-ref} is a fixed external baseline evaluated once. Rows are sorted by macro $r$ within each block, and bold marks the column-best value across the table.}
\centering
\footnotesize
\setlength{\tabcolsep}{4pt}
\renewcommand{\arraystretch}{0.9}
\begin{tabular}{lcccccccc}
\toprule
Method & sweet & bitter & salty & sour & spicy & Macro $r$ $\uparrow$ & MAE $\downarrow$ & RMSE $\downarrow$ \\
\midrule
MULE \cite{mccallum2022mule}                  & +0.831 & +0.645 & +0.609 & +0.610 & +0.641 & 0.667          & 0.116          & 0.146 \\
VGGish \cite{hershey2017vggish}               & \textbf{+0.870} & +0.728 & +0.367 & +0.578 & +0.787 & 0.666          & 0.109          & 0.134 \\
CLAP \cite{wu2023clap}                        & +0.798 & +0.607 & +0.651 & +0.598 & +0.666 & 0.664          & 0.114          & 0.142 \\
AST \cite{gong2021ast}                        & +0.824 & +0.709 & +0.580 & +0.615 & +0.564 & 0.658          & 0.116          & 0.143 \\
PANNs \cite{kong2020panns}                    & +0.816 & +0.608 & +0.533 & +0.528 & +0.673 & 0.632          & 0.119          & 0.146 \\
HuBERT \cite{hsu2021hubert}                   & +0.749 & +0.600 & +0.528 & +0.462 & +0.741 & 0.616          & 0.121          & 0.150 \\
EnCodec \cite{defossez2023encodec}            & +0.725 & +0.591 & +0.389 & +0.409 & +0.571 & 0.537          & 0.128          & 0.159 \\
MFCC (DSP floor)                              & +0.708 & +0.290 & +0.577 & +0.396 & +0.621 & 0.518          & 0.142          & 0.181 \\
MERT \cite{li2024mert}                        & +0.708 & +0.439 & +0.477 & +0.519 & +0.139 & 0.456          & 0.142          & 0.173 \\
Omar-RQ \cite{alonsojimenez2024omarrq}        & +0.450 & +0.523 & +0.562 & +0.438 & $-$0.357 & 0.323        & 0.141          & 0.169 \\
\midrule
\texttt{ast-5head-ref} (SOTA, \cite{rodriguez2024sonic}) & +0.740 & +0.563 & +0.468 & +0.325 & +0.682 & 0.556 & 0.175 & 0.219 \\
\midrule
\textbf{fusion VGGish$+$MULE}                 & +0.851 & \textbf{+0.756} & +0.643 & +0.657 & +0.715 & \textbf{0.724} & 0.111 & \textbf{0.134} \\
fusion CLAP$+$VGGish                          & +0.836 & +0.645 & +0.637 & +0.652 & \textbf{+0.788} & 0.712          & \textbf{0.109} & 0.135 \\
fusion AST$+$VGGish$+$MULE                    & +0.844 & +0.750 & +0.659 & \textbf{+0.697} & +0.600 & 0.710          & 0.112          & 0.137 \\
fusion CLAP$+$AST$+$VGGish$+$MULE             & +0.839 & +0.709 & +0.651 & +0.664 & +0.673 & 0.707          & 0.110          & 0.134 \\
fusion CLAP$+$MULE                            & +0.813 & +0.685 & \textbf{+0.665} & +0.648 & +0.696 & 0.701          & 0.114          & 0.139 \\
fusion AST$+$MULE                             & +0.844 & +0.728 & +0.652 & +0.647 & +0.580 & 0.690          & 0.113          & 0.140 \\
fusion AST$+$VGGish (gated)                   & +0.812 & +0.743 & +0.613 & +0.666 & +0.606 & 0.688          & 0.113          & 0.139 \\
\bottomrule
\end{tabular}
\label{tab:benchmark}
\end{table*}

\section{Results}

\subsection{Encoder benchmark and fusion}
\label{sec:results-benchmark}

Table~\ref{tab:benchmark} benchmarks ten frozen encoders under our shared multi-task head against the SOTA \texttt{ast-5head-ref}~\cite{rodriguez2024sonic} and seven late-fusion variants; every learned configuration is reported as a five-seed mean to keep the comparison stable on the $40$-item test set. We read macro RMSE and MAE as the primary metrics: the target use is to stand in for or augment human taste ratings, where what matters is how far a predicted rating sits from the true one in rating units, and the loss is masked MSE on $[0,1]$ targets. Macro Pearson $r$ is secondary, tracking rank quality for the content-based retrieval/indexing use case of \S\ref{sec:results-retrieval}. (i) Four encoders from three families tie at the top of the aggregate (MULE $r=0.667$, VGGish $0.666$, CLAP $0.664$, AST $0.658$, a $0.009$ spread within their seed noise), but their per-taste profiles differ: VGGish leads sweet, spicy, and bitter yet collapses on salty, where CLAP and MULE are most reliable and AST leads sour (Table~\ref{tab:benchmark}). No single family dominates across tastes, and this complementarity, also reported by~\cite{mccallum2022mule,frommholz2026predicting}, is what the fusion exploits. (ii) On absolute error the encoders are flat: a single \encoder{VGGish} already reaches the lowest RMSE ($0.134$) and MAE ($0.109$) in the table, which no fusion improves on, so for rating replacement one frozen encoder suffices. Gated late-fusion instead buys rank correlation: the strongest pair, \encoder{VGGish}$+$\encoder{MULE}, lifts macro $r$ to $0.724\pm0.020$ ($+0.057$ over \encoder{MULE}) at the same RMSE ($0.134$), with \encoder{CLAP}$+$\encoder{VGGish} ($0.712$) and the triple \encoder{AST}$+$\encoder{VGGish}$+$\encoder{MULE} ($0.710$) close behind. The seven fusions span only $0.688$--$0.724$ in macro $r$, mostly within one another's seed bands, so no pairing is decisively best, and per-target leadership stays spread across them (Table~\ref{tab:benchmark}), with the single \encoder{VGGish} still holding \emph{sweet} ($+0.870$). Stacking all four encoders does not help; complementarity matters more than count. (iii) Every learned configuration matches or beats the SOTA on macro RMSE (best $0.134$ vs.\ $0.219$), so the five-head fine-tuned design of~\cite{rodriguez2024sonic} carries a clear absolute-error cost.

\paragraph{Decomposing the gap.} Table~\ref{tab:ablation} attributes the gain over~\cite{rodriguez2024sonic} to three cumulative changes (five-seed means). The frozen-encoder $+$ per-taste-MLP $+$ masked-MSE $+$ sigmoid swap closes most of the RMSE gap ($0.219\!\to\!0.143$); collapsing the five heads to one shared multi-task head holds RMSE and $r$ ($0.663\!\to\!0.658$) at a $5\times$ smaller head; gated fusion adds the rest ($\to\!0.134$). Loss and output redesign therefore dominate over architecture (the first swap alone is $0.076$ of the $0.085$ RMSE gain), consistent with HEAR's finding that frozen-encoder benchmarks are sensitive to probe and loss~\cite{turian2022hear}; the gain transfers across AST, VGGish, MULE, and CLAP, so it is not a single-encoder artefact.

\begin{table}[t]
\caption{Cumulative ablation from the previous SOTA~\cite{rodriguez2024sonic} (row 1) to the best gated fusion (row 4). Each row stacks the change described in its leftmost column on top of the row above; rows 2--4 are five-seed means evaluated on the same held-out test split (row 4 fuses the best pair from Table~\ref{tab:benchmark}). Bold marks the column-best value.}
\centering
\footnotesize
\setlength{\tabcolsep}{4pt}
\resizebox{\columnwidth}{!}{%
\begin{tabular}{lccc}
\toprule
configuration & RMSE $\downarrow$ & MAE $\downarrow$ & macro $r$ $\uparrow$ \\
\midrule
SOTA \cite{rodriguez2024sonic} (fine-tuned AST, unbounded MSE)                              & 0.219          & 0.175          & 0.556 \\
\,+\,frozen AST, per-taste MLP, masked MSE, sigmoid                                          & 0.143          & 0.115          & 0.663 \\
\,+\,shared multi-task head (single AST encoder)                                            & 0.143          & 0.116          & 0.658 \\
\,+\,gated late-fusion VGGish$+$MULE                                                         & \textbf{0.134} & \textbf{0.111} & \textbf{0.724} \\
\bottomrule
\end{tabular}%
}
\label{tab:ablation}
\end{table}

\paragraph{Per-source breakdown.} The 40-item test set splits into 20 real-music~\cite{spanio2026multimodal} and 20 MusicGen-generated items~\cite{spanio2025multimodal} with different inter-rater agreement (Table~\ref{tab:source}). The single-rater leave-one-out ceiling on real music is $r_1\!\approx\!0.60$ and, in rating units, MAE $0.227$/RMSE $0.280$; the best systems' real-music error (MAE $\approx0.10$, RMSE $\approx0.13$) is less than half of that, so a frozen-encoder model estimates the group rating more closely than an average human rater, the core evidence that it can stand in for or be pooled with human ratings. The SOTA leads the real-music rank metrics ($r=0.767$, $\rho=0.751$), consistent with its soundtrack training, but AST$+$VGGish takes both error metrics there (MAE $0.099$, RMSE $0.130$), so its correlation edge does not extend to absolute error. On the harder generated subset ($r_1\!\approx\!0.25$) frozen AST leads three of four columns and VGGish$+$MULE takes RMSE ($0.133$), while SOTA's error is $\approx2.5\times$ ours: it underpredicts \emph{bitter}, \emph{salty}, and \emph{sour} by $0.25$--$0.37$ (its training corpus lacks MusicGen's high-intensity prompts~\cite{copet2024simple,spanio2025multimodal}), whereas the frozen-encoder systems see both sources at training. VGGish$+$MULE is the only system competitive on both subsets at once.

\begin{table}[t]
\caption{Source-aware metrics on the two test subsets ($n=20$ each): macro Pearson $r$, Spearman $\rho$, MAE, and RMSE. Bottom row: a single human rater's leave-one-out (LOO) agreement with the consensus, as Pearson $r$ (averaged over the per-taste values) and as absolute error. The real-music error ceiling ($0.227$/$0.280$) is computed from the 49-rater study; the generated subset has no comparable per-rater error estimate, and far lower agreement (Krippendorff $\alpha\approx0.1$--$0.2$, only $39\%$ of clips heard as their prompt~\cite{spanio2025multimodal}). Bold marks the column-best value among the models.}
\centering
\footnotesize
\setlength{\tabcolsep}{2.5pt}
\resizebox{\columnwidth}{!}{%
\begin{tabular}{l cccc | cccc}
\toprule
                                & \multicolumn{4}{c|}{real-music ($n{=}20$)} & \multicolumn{4}{c}{MusicGen-generated ($n{=}20$)} \\
method                          & $r$ $\uparrow$ & $\rho$ $\uparrow$ & MAE $\downarrow$ & RMSE $\downarrow$ & $r$ $\uparrow$ & $\rho$ $\uparrow$ & MAE $\downarrow$ & RMSE $\downarrow$ \\
\midrule
\encoder{AST}                    & 0.613 & 0.645 & 0.112 & 0.137 & \textbf{0.559} & \textbf{0.581} & \textbf{0.109} & 0.135 \\
\encoder{VGGish}                 & 0.583 & 0.605 & 0.110 & 0.132 & 0.486 & 0.482 & 0.116 & 0.144 \\
\encoder{MULE}                   & 0.650 & 0.653 & 0.116 & 0.145 & 0.433 & 0.420 & 0.122 & 0.149 \\
\encoder{CLAP}                   & 0.659 & 0.678 & 0.115 & 0.140 & 0.504 & 0.519 & 0.118 & 0.145 \\
SOTA \cite{rodriguez2024sonic}   & \textbf{0.767} & \textbf{0.751} & 0.112 & 0.142 & 0.454 & 0.491 & 0.262 & 0.295 \\
pair AST$+$VGGish                & 0.662 & 0.674 & \textbf{0.099} & \textbf{0.130} & 0.537 & 0.555 & 0.115 & 0.143 \\
pair VGGish$+$MULE               & 0.694 & 0.689 & 0.111 & 0.132 & 0.539 & 0.564 & 0.110 & \textbf{0.133} \\
triple AST$+$VGGish$+$MULE       & 0.678 & 0.691 & 0.116 & 0.138 & 0.429 & 0.420 & 0.119 & 0.147 \\
\midrule
single human rater (LOO)         & $\approx 0.60$ & --- & 0.227 & 0.280 & $\approx 0.25$ & --- & --- & --- \\
\bottomrule
\end{tabular}%
}
\label{tab:source}
\end{table}

\paragraph{Robustness and the small-$N$ caveat.} Seed spread is small (on AST, five-seed $r=0.658\pm0.010$, RMSE $0.143\pm0.003$) but dwarfed by sampling uncertainty on the $40$-item test set: a $2000$-resample bootstrap gives the best fusion's macro $r$ a $95\%$ CI roughly $0.18$ wide ($[0.64,0.82]$). The seven fusions therefore lie inside one another's intervals, so their ordering is indicative, not significant; only the large gaps clear a paired bootstrap at $p<0.05$ (any learned encoder over the SOTA, fusion over a single encoder). The findings are stable to the head: linear ridge, $k$NN, heteroscedastic, and Gaussian-process heads stay within $\pm0.05$ macro $r$ of the MLP, the ridge probe nearly matching it, consistent with a largely linear audio--taste relationship.

\subsection{Psychophysics-grounded analysis explains why the strongest models work}
\label{sec:results-probes}

\begin{figure*}[!t]
\centering
\includegraphics[width=0.82\textwidth]{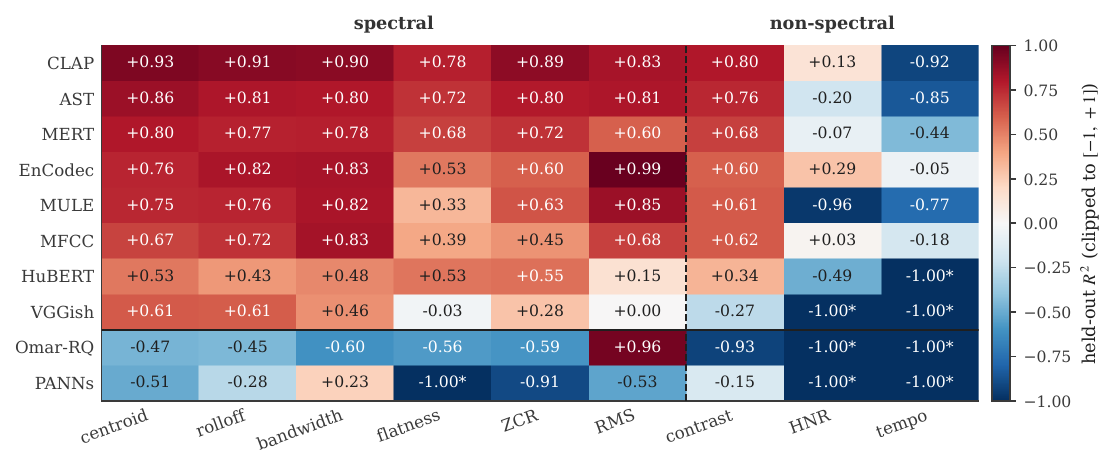}
\caption{Held-out $R^2$ of $5$-fold ridge probes from each cached embedding to nine psychoacoustic descriptors. Encoders are sorted by mean spectral $R^2$; the horizontal line separates positive- from negative-mean encoders, and the dashed vertical line separates spectral (left) from non-spectral (right) features. The color scale is clipped to $[-1, +1]$; cells whose raw $R^2$ falls outside this window are marked with~$\ast$.}
\label{fig:probes}
\end{figure*}

\paragraph{Psychoacoustic probes: spectral structure predicts taste only loosely.} Fig.~\ref{fig:probes} reports held-out $R^2$ for ridge probes from each embedding to nine psychoacoustic descriptors, the cue family the crossmodal literature ties to sound--taste associations~\cite{spence2011crossmodal,knoferle2012sounds,wang2017assessing}. CLAP leads \emph{every} spectral column ($R^2\!\in\![0.78,0.93]$), so a contrastive audio--language objective does not discard fine spectral structure; AST follows, then MERT, EnCodec, and MULE, strong on the brightness and bandwidth axes psychophysics ties to sweet and bitter~\cite{crisinel2010crossmodal,watson2017multisensory}. But decoding spectra and predicting taste track each other only loosely in both directions: MERT and EnCodec probe well yet rank near the bottom of Table~\ref{tab:benchmark}, while VGGish and PANNs predict taste well despite weak spectral $R^2$. The encoders reach taste through partly-independent representations, which is what their gated fusion exploits; Omar-RQ and PANNs sit far below zero on most spectral columns, mirroring~\cite{frommholz2026predicting}. Loudness stays trivially decodable from a codec objective (EnCodec RMS $R^2=0.987$), and tempo is uniformly weak, consistent with it being the noisiest crossmodal axis~\cite{spence2017gastrophysics,guedes2023crossmodal}.

\paragraph{Bandstop knockout exposes different mechanisms.} A second probe applies a zero-phase 4th-order Butterworth bandstop (forward--backward, RMS-renormalised) at each of eight mel-scale bands and measures the per-(band, taste) Pearson drop $\Delta r$ after re-encoding (Table~\ref{tab:mechanism}).

\begin{table}[t]
\caption{Cross-encoder audio-bandstop knockout for the four strongest encoders. For each taste, the input frequency band whose removal most changes per-taste Pearson $r$ on the held-out test split. Negative $\Delta r$: band removal hurts performance; positive (band ``\emph{none}''): the most-affected band was a nuisance. Bold marks the most-damaging band removal per encoder; CLAP has none ($|\Delta r| \le 0.064$ everywhere).}
\centering
\footnotesize
\setlength{\tabcolsep}{2.5pt}
\resizebox{\columnwidth}{!}{%
\begin{tabular}{l l r l r l r l r}
\toprule
taste & AST & $\Delta r$ & VGGish & $\Delta r$ & CLAP & $\Delta r$ & MULE & $\Delta r$ \\
\midrule
sweet  & 260--610\,Hz & $-0.039$          & 260--610\,Hz  & $-0.060$           & 260--610\,Hz & $-0.038$ & 260--610\,Hz  & $-0.125$ \\
bitter & 0--260\,Hz   & $\mathbf{-0.129}$ & 0--260\,Hz    & $\mathbf{-0.166}$  & \emph{none}  & $+0.038$ & 3.9--5.6\,kHz & $-0.090$ \\
salty  & \emph{none}  & $+0.075$          & 3.9--5.6\,kHz & $-0.116$           & 0--260\,Hz   & $-0.064$ & 1.1--1.8\,kHz & $\mathbf{-0.294}$ \\
sour   & \emph{none}  & $+0.059$          & 3.9--5.6\,kHz & $-0.143$           & \emph{none}  & $+0.043$ & 1.8--2.7\,kHz & $-0.149$ \\
spicy  & \emph{none}  & $+0.101$          & \emph{none}   & $+0.066$           & 260--610\,Hz & $-0.024$ & 1.8--2.7\,kHz & $-0.264$ \\
\bottomrule
\end{tabular}%
}
\label{tab:mechanism}
\end{table}

The four encoders rely on different mechanisms. \encoder{AST} concentrates per-taste signal in one or two narrow bands (sweet at 260--610\,Hz, bitter at sub-bass), with positive $\Delta r$ on salty, sour, and spicy (band-localised nuisance detectors, e.g.\ an anti-bass filter for spicy); \encoder{VGGish} splits its dependencies between sub-bass (bitter, $-0.166$) and $3.9$--$5.6$\,kHz upper-mids (salty, sour); \encoder{CLAP} is band-robust (no removal exceeds $|0.064|$), the strongest spectral decoder in Fig.~\ref{fig:probes}; \encoder{MULE} is uniformly negative, concentrated in $1$--$3$\,kHz (salty $-0.294$, spicy $-0.264$). The macro-$r$ benchmark hides this: the headline \encoder{VGGish}$+$\encoder{MULE} pair combines two negative-dependence encoders whose critical bands are disjoint (sub-bass$+$upper-mids vs.\ $1$--$3$\,kHz), exactly the complementarity its gating exploits. The mappings match the literature: bitter-to-sub-bass tracks the replicated finding that low-pitched dark sounds elicit more bitterness~\cite{knoferle2012sounds,watson2017multisensory,mesz2012composition}, sweet-to-$260$--$610$\,Hz sits in the bright-formant range reported as crossmodally sweet~\cite{crisinel2010crossmodal,wang2016multisensory,knoferle2015pm}, and \encoder{MULE}'s $1$--$3$\,kHz reliance overlaps the salty/sour region of~\cite{wang2021multisensory,qi2020perception}. The bandstop is a blunt probe~\cite{frommholz2026predicting}, but the split is consistent across tastes and robust to filter-order perturbations.

\subsection{Retrieval by taste profile}
\label{sec:results-retrieval}

This is the second use case for the taste space, and the one closest to this venue's content-based indexing focus: organising a collection by predicted taste, where rank fidelity matters rather than absolute error, so Spearman $\rho$ and Pearson $r$ are the metrics of record. We index test items by their predicted 5-D vectors and rank them by Euclidean distance to five taste-profile queries (\texttt{dessert}, \texttt{dark-chocolate}, \texttt{umami-savory}, \texttt{citrus}, \texttt{chili-burn}), mirroring~\cite{wilkins2026controllable}'s attribute-guided setup for mood. Each query is the mean 5-D taste vector of foods with that dominant taste in the dataset of~\cite{spanio2026multimodal}, preserving real-food structure (\texttt{dessert} carries sour from fruit; \texttt{dark-chocolate} pairs bitter and sweet) rather than collapsing to one-hot.

\paragraph{Why the $309$-item pool and three metrics.} On the 40-item test set alone, median-binarised $P@k$ saturates ($>0.9$ at $k=5$) for every reasonable predictor and cannot rank systems, a known small-pool limitation~\cite{wilkins2026controllable}. We add the 269 training items as in-distribution distractors and report $P@k$ on the 309-pool, test-only Spearman $\rho$ on the 40 labeled items, and test-vs-distractor ROC-AUC. Neither $\rho$ nor AUC uses distractor labels, so the evaluation is not confounded by SOTA's weak-label training data~\cite{rodriguez2024sonic} (random baselines: $P@5\,0.50$, AUC $0.5$, $\rho\,0$).

\begin{table}[t]
\caption{Retrieval-by-taste-profile on a $309$-item pool ($40$ test items + $269$ training items as in-distribution distractors), mean across the five food-profile queries derived from~\cite{spanio2026multimodal}. $P@k$: precision at $k$ under median-binarised ground-truth distance to the query. $\rho$: Spearman rank correlation between predicted and ground-truth distance on the $40$ labeled test items. AUC: ROC-AUC of test-vs-distractor discrimination ($0.5$ = chance). CLAP-text~\cite{wu2023clap} ranks audio by cosine similarity to a textual paraphrase of each query; it surfaces a fully-observed item in the top-$k$ for at most one query, so its $P@k$ is undefined (``---'') and its load-bearing $\rho$/AUC are reported instead. Bold marks the column-best value.}
\centering
\small
\setlength{\tabcolsep}{4pt}
\resizebox{\columnwidth}{!}{%
\begin{tabular}{l ccccc}
\toprule
system                                              & $P@5$ & $P@10$ & $P@20$ & $\rho$ (test) & AUC \\
\midrule
pair AST$+$VGGish                                    & \textbf{1.000} & \textbf{1.000} & \textbf{1.000} & \textbf{0.693} & 0.468 \\
pair VGGish$+$MULE                                   & \textbf{1.000} & \textbf{1.000} & \textbf{1.000} & 0.663          & 0.478 \\
triple AST$+$VGGish$+$MULE                           & \textbf{1.000} & 0.833          & 0.875          & 0.663          & 0.510 \\
SOTA \cite{rodriguez2024sonic}                       & \textbf{1.000} & \textbf{1.000} & \textbf{1.000} & 0.645          & 0.477 \\
\midrule
CLAP-text \cite{wu2023clap}                          & ---            & ---            & ---            & 0.122          & 0.484 \\
random (chance)                                      & 0.500          & 0.500          & 0.500          & 0.000          & 0.500 \\
\bottomrule
\end{tabular}%
}
\label{tab:retrieval}
\end{table}

Table~\ref{tab:retrieval} yields two findings. First, taste-space retrieval far outperforms text-space: every 5-D predictor saturates $P@k$ wherever defined, whereas CLAP-text's $\rho$ ($0.122$) and AUC ($0.484$) sit at chance, surfacing a taste-relevant item for only one of five queries, replicating the text-vs-attribute gap of~\cite{wilkins2026controllable}. Second, the regression-best fusion is not the rank-best: since precision cannot separate the predictors we read $\rho$, on which \encoder{AST}$+$\encoder{VGGish} ranks most faithfully ($0.693$), edging \encoder{VGGish}$+$\encoder{MULE} and the triple (both $0.663$) and SOTA ($0.645$~\cite{rodriguez2024sonic}). So the model to deploy depends on the task: a single \encoder{VGGish} for low-error rating, \encoder{AST}$+$\encoder{VGGish} for indexing.

\paragraph{Per-query variance.} Per-query $\rho$ for the best fusion spans $0.474$ (\texttt{citrus}) to $0.820$ (\texttt{dark-chocolate}), highest on sweet/bitter-dominant queries, consistent with their larger crossmodal effects~\cite{spence2011crossmodal,knoferle2012sounds,guedes2023crossmodal}. AUC is near chance by construction, since the distractors share the test items'~\cite{spanio2026multimodal} predicted-taste distribution, so the test-only $\rho$ is the load-bearing metric, haystack-invariant when the distractors are swapped for $\sim\!5{,}000$ OOD FMA chunks~\cite{defferrard2017fma}.

\section{Conclusion}
\label{sec:conclusion}

We formalised taste-from-audio prediction as a content-based MIR benchmark over a perceptually validated corpus~\cite{spanio2026multimodal}: a frozen-encoder protocol over ten encoders across the four HEAR families~\cite{turian2022hear}, source-aware reporting against inter-rater ceilings, retrieval over a $309$-item pool, and a psychophysics-grounded interpretability layer. Read for rating replacement, the best systems predict the five tastes within a macro RMSE of $0.134$; on held-out real music their error is less than half a single rater's deviation from the consensus (RMSE $0.13$ vs.\ $0.28$), so a frozen-encoder model estimates the group rating more closely than an average human, and a single \encoder{VGGish} reaches it. The ablation (Table~\ref{tab:ablation}) shows this gain over the prior baseline comes from loss and output redesign, not architecture; gated late-fusion adds only rank correlation (macro $r$ $0.724$ vs.\ $0.666$), useful for the indexing use case where a generic CLAP-text~\cite{wu2023clap} baseline collapses to chance. Because the predicted taste vector has a natural-language reading, it can supply content-based recommendation~\cite{vandenoord2013deep,schedl2018current} with an interpretable axis.

\textbf{Limitations.} The evaluation is small ($269$ training, $40$ test clips, $n=20$ per source), which bounds generalisability and favours simple heads (the ridge probe nearly matches the MLP); scaling the annotated corpus, not the model, is the main lever. The crossmodal literature we lean on is Western and sweet/bitter-centric, so cross-cultural generalisation~\cite{spanio2025multimodal} and the under-constrained spicy/trigeminal axis~\cite{guedes2023crossmodal} are open. Code and trained heads are at \texttt{\url{https://github.com/CSCPadova/wav2taste}}; the corpus is public at \texttt{\url{https://huggingface.co/datasets/csc-unipd/sonic-seasoning}}.

\bibliographystyle{ieeetr}
\bibliography{references}

@article{spanio2025multimodal,
AUTHOR={Spanio, Matteo  and Zampini, Massimiliano  and Rodà, Antonio  and Pierucci, Franco },
TITLE={A multimodal symphony: integrating taste and sound through generative AI},
JOURNAL={Frontiers in Computer Science},
VOLUME={Volume 7 - 2025},
YEAR={2025},
URL={https://www.frontiersin.org/journals/computer-science/articles/10.3389/fcomp.2025.1575741},
DOI={10.3389/fcomp.2025.1575741},
ISSN={2624-9898}}

@misc{spanio2026multimodal,
      title={Multimodal Dataset Normalization and Perceptual Validation for Music-Taste Correspondences}, 
      author={Matteo Spanio and Valentina Frezzato and Antonio Rodà},
      year={2026},
      eprint={2604.10632},
      archivePrefix={arXiv},
      primaryClass={cs.SD},
      url={https://arxiv.org/abs/2604.10632}, 
}

@article{guedes2023taste,
  author    = {David Guedes and Marília Prada and Margarida Vaz Garrido and Elsa Lamy},
  title     = {The Taste \& Affect Music Database: Subjective Rating Norms for a New Set of Musical Stimuli},
  journal   = {Behavior Research Methods},
  year      = {2023},
  volume    = {55},
  number    = {3},
  pages     = {1121--1140},
  doi       = {10.3758/s13428-022-01862-z},
  url       = {https://doi.org/10.3758/s13428-022-01862-z},
  issn      = {1554-3528}
}

@article{spence2011crossmodal,
  author    = {Charles Spence},
  title     = {Crossmodal Correspondences: A Tutorial Review},
  journal   = {Attention, Perception, \& Psychophysics},
  year      = {2011},
  volume    = {73},
  number    = {4},
  pages     = {971--995},
  doi       = {10.3758/s13414-010-0073-7},
  url       = {https://doi.org/10.3758/s13414-010-0073-7},
  issn      = {1943-393X}
}

@article{knoferle2012sounds,
  author    = {Klemens Kn{\"o}ferle and Charles Spence},
  title     = {Crossmodal Correspondences Between Sounds and Tastes},
  journal   = {Psychonomic Bulletin \& Review},
  year      = {2012},
  volume    = {19},
  number    = {6},
  pages     = {992--1006},
  doi       = {10.3758/s13423-012-0321-z},
  url       = {https://doi.org/10.3758/s13423-012-0321-z},
  issn      = {1531-5320}
}

@article{wang2017assessing,
author = {Wang, Qian (Janice) and Spence, Charles},
title = {Assessing the influence of music on wine perception among wine professionals},
journal = {Food Science \& Nutrition},
volume = {6},
number = {2},
pages = {295-301},
doi = {https://doi.org/10.1002/fsn3.554},
url = {https://onlinelibrary.wiley.com/doi/abs/10.1002/fsn3.554},
eprint = {https://onlinelibrary.wiley.com/doi/pdf/10.1002/fsn3.554},
year = {2018}
}

@book{spence2017gastrophysics,
  title={Gastrophysics: The New Science of Eating},
  author={Spence, C.},
  isbn={9780241977736},
  url={https://books.google.it/books?id=D_E0DQAAQBAJ},
  year={2017},
  publisher={Penguin Books Limited}
}

@INPROCEEDINGS{hershey2017vggish,
  author={Hershey, Shawn and Chaudhuri, Sourish and Ellis, Daniel P. W. and Gemmeke, Jort F. and Jansen, Aren and Moore, R. Channing and Plakal, Manoj and Platt, Devin and Saurous, Rif A. and Seybold, Bryan and Slaney, Malcolm and Weiss, Ron J. and Wilson, Kevin},
  booktitle={2017 IEEE International Conference on Acoustics, Speech and Signal Processing (ICASSP)}, 
  title={CNN architectures for large-scale audio classification}, 
  year={2017},
  volume={},
  number={},
  pages={131-135},
  doi={10.1109/ICASSP.2017.7952132}}

@ARTICLE{kong2020panns,
  author={Kong, Qiuqiang and Cao, Yin and Iqbal, Turab and Wang, Yuxuan and Wang, Wenwu and Plumbley, Mark D.},
  journal={IEEE/ACM Transactions on Audio, Speech, and Language Processing}, 
  title={PANNs: Large-Scale Pretrained Audio Neural Networks for Audio Pattern Recognition}, 
  year={2020},
  volume={28},
  number={},
  pages={2880-2894},
  doi={10.1109/TASLP.2020.3030497}}

@inproceedings{gong2021ast,
  title     = {{AST: Audio Spectrogram Transformer}},
  author    = {Yuan Gong and Yu-An Chung and James Glass},
  year      = {2021},
  booktitle = {{Interspeech 2021}},
  pages     = {571--575},
  doi       = {10.21437/Interspeech.2021-698},
  issn      = {2958-1796},
}

@ARTICLE{hsu2021hubert,
  author={Hsu, Wei-Ning and Bolte, Benjamin and Tsai, Yao-Hung Hubert and Lakhotia, Kushal and Salakhutdinov, Ruslan and Mohamed, Abdelrahman},
  journal={IEEE/ACM Transactions on Audio, Speech, and Language Processing}, 
  title={HuBERT: Self-Supervised Speech Representation Learning by Masked Prediction of Hidden Units}, 
  year={2021},
  volume={29},
  number={},
  pages={3451-3460},
  doi={10.1109/TASLP.2021.3122291}}

@INPROCEEDINGS{wu2023clap,
  author={Wu, Yusong and Chen, Ke and Zhang, Tianyu and Hui, Yuchen and Berg-Kirkpatrick, Taylor and Dubnov, Shlomo},
  booktitle={ICASSP 2023 - 2023 IEEE International Conference on Acoustics, Speech and Signal Processing (ICASSP)}, 
  title={Large-Scale Contrastive Language-Audio Pretraining with Feature Fusion and Keyword-to-Caption Augmentation}, 
  year={2023},
  volume={},
  number={},
  pages={1-5},
  doi={10.1109/ICASSP49357.2023.10095969}}

@article{defossez2023encodec,
title={High Fidelity Neural Audio Compression},
author={Alexandre D{\'e}fossez and Jade Copet and Gabriel Synnaeve and Yossi Adi},
journal={Transactions on Machine Learning Research},
issn={2835-8856},
year={2023},
url={https://openreview.net/forum?id=ivCd8z8zR2},
}

@inproceedings{li2024mert,
  author    = {Li, Yizhi and Yuan, Ruibin and Zhang, Ge and Ma, Yinghao and Chen, Xingran and Yin, Hanzhi and Lin, Chenghao and Ragni, Anton and Benetos, Emmanouil and Gyenge, Norbert and Dannenberg, Roger B. and Liu, Ruibo and Chen, Wenhu and Xia, Gus and Shi, Yemin and Huang, Wenhao and Wang, Zili and Guo, Yike and Fu, Jie},
  title     = {{MERT}: Acoustic Music Understanding Model with Large-Scale Self-Supervised Training},
  booktitle = {International Conference on Learning Representations},
  year      = {2024},
  url       = {https://openreview.net/forum?id=w3YZ9MSlBu}
}

@inproceedings{mccallum2022mule,
  author    = {McCallum, Matthew C. and Korzeniowski, Filip and Oramas, Sergio and Gouyon, Fabien and Ehmann, Andreas F.},
  title     = {Supervised and Unsupervised Learning of Audio Representations for Music Understanding},
  booktitle = {Proceedings of the International Society for Music Information Retrieval Conference},
  year      = {2022}
}

@INPROCEEDINGS{wilkins2026controllable,
  author={Wilkins, Julia and Kim, Jaehun and Davies, Matthew E. P. and Pablo Bello, Juan and McCallum, Matthew C.},
  booktitle={ICASSP 2026 - 2026 IEEE International Conference on Acoustics, Speech and Signal Processing (ICASSP)}, 
  title={Controllable Embedding Transformation for Mood-Guided Music Retrieval}, 
  year={2026},
  volume={},
  number={},
  pages={15262-15266},
  doi={10.1109/ICASSP55912.2026.11461460}}

@inproceedings{spanio2024towards, author = {Spanio, Matteo}, language = {en}, title = {Towards Emotionally Aware AI: Challenges and Opportunities in the Evolution of Multimodal Generative Models}, booktitle = {Proceedings of the AIxIA Doctoral Consortium 2024 co-located with the 23nd International Conference of the Italian Association for Artificial Intelligence (AIxIA 2024)}, series = {{CEUR} Workshop Proceedings}, year = {2024}, publisher = {http://CEUR-WS.org}, url = {https://ceur-ws.org/Vol-3914/}, }

@phdthesis{rodriguez2024sonic,
  title={Listening to sustainable bites: Assessing the influence of sound on sustainable food perceptions and behaviors using a data-driven approach},
  author={Rodr{\'\i}guez Rivera, Brayan Mauricio},
  year={2025},
  month={June},
  school={Universidad de los Andes},
  type={PhD thesis},
  address={Bogotá - Colombia}
}

@article{mcfee2015,
  author = {McFee, Brian and Raffel, Colin and Liang, Dawen and Ellis, Daniel P.W. and McVicar, Matt and Battenberg, Eric and Nieto, Oriol},
  title = {librosa: Audio and Music Signal Analysis in Python},
  journal = {SciPy 2015},
  year = {2015},
  doi = {10.25080/Majora-7b98e3ed-003},
  url = {https://doi.org/10.25080/Majora-7b98e3ed-003}
}

@article{crisinel2010crossmodal,
author = {Anne-Sylvie Crisinel and Charles Spence},
title ={A Sweet Sound? Food Names Reveal Implicit Associations between Taste and Pitch},
journal = {Perception},
volume = {39},
number = {3},
pages = {417-425},
year = {2010},
doi = {10.1068/p6574},
    note ={PMID: 20465176},
URL = { 
        https://doi.org/10.1068/p6574
},
eprint = { 
        https://doi.org/10.1068/p6574
}
}

@inproceedings{alonsojimenez2024omarrq,
  title = {{OMAR-RQ}: Open Music Audio Representation Model Trained with Multi-Feature Masked Token Prediction},
  author = {Alonso-Jim{\'e}nez, Pablo and Ramoneda, Pedro and Araz, R. Oguz and Poltronieri, Andrea and Bogdanov, Dmitry},
  booktitle = {ACM Multimedia Conference (ACMMM), Open Source Track},
  year = {2025},
  doi = {10.1145/3746027.3756871},
}

@article{frommholz2026predicting,
 author = {Frommholz, Annika and Lepa, Steffen and Virkus, Tom and Weinzierl, Stefan  and Helberger, Johannes},
 doi = {10.5334/tismir.290},
 journal = {Transactions of the International Society for Music Information Retrieval},
 month = {Mar},
 title = {Predicting Perceived Semantic Expression of Functional Sounds Using Unsupervised Feature Extraction and Ensemble Learning},
 year = {2026}
}

@InProceedings{kornblith2019cka,
  title = 	 {Similarity of Neural Network Representations Revisited},
  author =       {Kornblith, Simon and Norouzi, Mohammad and Lee, Honglak and Hinton, Geoffrey},
  booktitle = 	 {Proceedings of the 36th International Conference on Machine Learning (ICML)},
  pages = 	 {3519--3529},
  year = 	 {2019},
  editor = 	 {Chaudhuri, Kamalika and Salakhutdinov, Ruslan},
  volume = 	 {97},
  series = 	 {Proceedings of Machine Learning Research},
  month = 	 {09--15 Jun},
  publisher =    {PMLR},
  url = 	 {https://proceedings.mlr.press/v97/kornblith19a.html}
}

@InProceedings{turian2022hear,
  title = 	 {{HEAR: Holistic Evaluation of Audio Representations}},
  author =       {Turian, Joseph and Shier, Jordie and Khan, Humair Raj and Raj, Bhiksha and Schuller, Bj\"{o}rn W. and Steinmetz, Christian J. and Malloy, Colin and Tzanetakis, George and Velarde, Gissel and McNally, Kirk and Henry, Max and Pinto, Nicolas and Noufi, Camille and Clough, Christian and Herremans, Dorien and Fonseca, Eduardo and Engel, Jesse and Salamon, Justin and Esling, Philippe and Manocha, Pranay and Watanabe, Shinji and Jin, Zeyu and Bisk, Yonatan},
  booktitle = 	 {Proceedings of the NeurIPS 2021 Competitions and Demonstrations Track},
  pages = 	 {125--145},
  year = 	 {2022},
  editor = 	 {Kiela, Douwe and Ciccone, Marco and Caputo, Barbara},
  volume = 	 {176},
  series = 	 {Proceedings of Machine Learning Research},
  month = 	 {06--14 Dec},
  publisher =    {PMLR},
  url = 	 {https://proceedings.mlr.press/v176/turian22a.html}
}

@article{knoferle2015pm,
author = {Knoeferle, Klemens M. and Woods, Andy and Käppler, Florian and Spence, Charles},
title = {That Sounds Sweet: Using Cross-Modal Correspondences to Communicate Gustatory Attributes},
journal = {Psychology \& Marketing},
volume = {32},
number = {1},
pages = {107-120},
doi = {https://doi.org/10.1002/mar.20766},
url = {https://onlinelibrary.wiley.com/doi/abs/10.1002/mar.20766},
eprint = {https://onlinelibrary.wiley.com/doi/pdf/10.1002/mar.20766},
year = {2015}
}

@article{wang2016multisensory,
    author = {Wang, Qian Janice and Wang, Sheila and Spence, Charles},
    title = {“Turn Up the Taste”: Assessing the Role of Taste Intensity and Emotion in Mediating Crossmodal Correspondences between Basic Tastes and Pitch},
    journal = {Chemical Senses},
    volume = {41},
    number = {4},
    pages = {345-356},
    year = {2016},
    month = {05},
    issn = {0379-864X},
    doi = {10.1093/chemse/bjw007},
    url = {https://doi.org/10.1093/chemse/bjw007},
    eprint = {https://academic.oup.com/chemse/article-pdf/41/4/345/17428060/bjw007.pdf},
}

@inproceedings{wang2021multisensory,
author = {Wang, Qian Janice and Mesz, Bruno and Spence, Charles},
title = {Assessing the impact of music on basic taste perception using time intensity analysis},
year = {2017},
isbn = {9781450355568},
publisher = {Association for Computing Machinery},
address = {New York, NY, USA},
url = {https://doi.org/10.1145/3141788.3141792},
doi = {10.1145/3141788.3141792},
booktitle = {Proceedings of the 2nd ACM SIGCHI International Workshop on Multisensory Approaches to Human-Food Interaction},
pages = {18–22},
numpages = {5},
location = {Glasgow, UK},
series = {MHFI 2017}
}

@article{qi2020perception,
author = {Barbosa Escobar, Francisco and Wang, Qian Janice},
title = {Inducing Novel Sound–Taste Correspondences via an Associative Learning Task},
journal = {Cognitive Science},
volume = {48},
number = {3},
pages = {e13421},
doi = {https://doi.org/10.1111/cogs.13421},
url = {https://onlinelibrary.wiley.com/doi/abs/10.1111/cogs.13421},
eprint = {https://onlinelibrary.wiley.com/doi/pdf/10.1111/cogs.13421},
year = {2024}
}

@ARTICLE{mesz2012composition,
    
AUTHOR={Mesz, Bruno  and Sigman, Mariano  and Trevisan, Marcos },
           
TITLE={A composition algorithm based on crossmodal taste-music correspondences},
          
JOURNAL={Frontiers in Human Neuroscience},
          
VOLUME={Volume 6 - 2012},
  
YEAR={2012},
  
URL={https://www.frontiersin.org/journals/human-neuroscience/articles/10.3389/fnhum.2012.00071},
  
DOI={10.3389/fnhum.2012.00071},
  
ISSN={1662-5161}}

@article{guedes2023crossmodal,
title = {Crossmodal interactions between audition and taste: A systematic review and narrative synthesis},
journal = {Food Quality and Preference},
volume = {107},
pages = {104856},
year = {2023},
issn = {0950-3293},
doi = {https://doi.org/10.1016/j.foodqual.2023.104856},
url = {https://www.sciencedirect.com/science/article/pii/S0950329323000502},
author = {David Guedes and Margarida {Vaz Garrido} and Elsa Lamy and Bernardo {Pereira Cavalheiro} and Marília Prada}
}

@article{mathiesen2022sound,
title = {Harmonising flavours: How arousing music and sound influence food perception and emotional responses},
journal = {International Journal of Gastronomy and Food Science},
volume = {39},
pages = {101093},
year = {2025},
issn = {1878-450X},
doi = {https://doi.org/10.1016/j.ijgfs.2024.101093},
url = {https://www.sciencedirect.com/science/article/pii/S1878450X24002269},
author = {Yi Hsuan Tiffany Lin and Daniel Shepherd and Kevin Kantono and Charles Spence and Nazimah Hamid}
}

@article{galmarini2021impact,
title = {Impact of music on the dynamic perception of coffee and evoked emotions evaluated by temporal dominance of sensations (TDS) and emotions (TDE)},
journal = {Food Research International},
volume = {150},
pages = {110795},
year = {2021},
issn = {0963-9969},
doi = {https://doi.org/10.1016/j.foodres.2021.110795},
url = {https://www.sciencedirect.com/science/article/pii/S0963996921006955},
author = {M.V. Galmarini and R.J. {Silva Paz} and D. {Enciso Choquehuanca} and M.C. Zamora and B. Mesz}
}

@article{watson2017multisensory,
  author    = {Quentin J. Watson and Karen L. Gunther},
  title     = {Trombones Elicit Bitter More Strongly Than Do Clarinets: A Partial Replication of Three Studies of Crisinel and Spence},
  journal   = {Multisensory Research},
  year      = {2017},
  volume    = {30},
  number    = {3--5},
  pages     = {321--335},
  doi       = {10.1163/22134808-00002573},
  issn      = {2213-4794},
  eissn     = {2213-4808}
}

@article{zampini2010sound,
  author    = {Massimiliano Zampini and Charles Spence},
  title     = {Assessing the Role of Sound in the Perception of Food and Drink},
  journal   = {Chemosensory Perception},
  year      = {2010},
  volume    = {3},
  number    = {1},
  pages     = {57--67},
  doi       = {10.1007/s12078-010-9064-2},
  url       = {https://doi.org/10.1007/s12078-010-9064-2},
  issn      = {1936-5810}
}

@inproceedings{copet2024simple,
 author = {Copet, Jade and Kreuk, Felix and Gat, Itai and Remez, Tal and Kant, David and Synnaeve, Gabriel and Adi, Yossi and Defossez, Alexandre},
 booktitle = {Advances in Neural Information Processing Systems},
 editor = {A. Oh and T. Naumann and A. Globerson and K. Saenko and M. Hardt and S. Levine},
 pages = {47704--47720},
 publisher = {Curran Associates, Inc.},
 title = {Simple and Controllable Music Generation},
 url = {https://proceedings.neurips.cc/paper_files/paper/2023/file/94b472a1842cd7c56dcb125fb2765fbd-Paper-Conference.pdf},
 volume = {36},
 year = {2023}
}

@inproceedings{defferrard2017fma,
  title = {{FMA}: A Dataset for Music Analysis},
  author = {Defferrard, Micha\"el and Benzi, Kirell and Vandergheynst, Pierre and Bresson, Xavier},
  booktitle = {18th International Society for Music Information Retrieval Conference (ISMIR)},
  year = {2017},
  archiveprefix = {arXiv},
  eprint = {1612.01840},
  url = {https://arxiv.org/abs/1612.01840},
}

@article{lepa2020computational,
author = {Steffen Lepa and Martin Herzog and Jochen Steffens and Andreas Schoenrock and Hauke Egermann},
title = {A computational model for predicting perceived musical expression in branding scenarios},
journal = {Journal of New Music Research},
volume = {49},
number = {4},
pages = {387--402},
year = {2020},
publisher = {Routledge},
doi = {10.1080/09298215.2020.1778041},
URL = { 
        https://doi.org/10.1080/09298215.2020.1778041
},
eprint = { 
        https://doi.org/10.1080/09298215.2020.1778041
}
}

@ARTICLE{kang2024music,
  author={Kang, Jaeyong and Herremans, Dorien},
  journal={IEEE Transactions on Affective Computing}, 
  title={Are We There Yet? A Brief Survey of Music Emotion Prediction Datasets, Models and Outstanding Challenges}, 
  year={2025},
  volume={16},
  number={4},
  pages={2545-2559},
  doi={10.1109/TAFFC.2025.3583505}}

@inproceedings{vandenoord2013deep,
author = {Oord, A\"{a}ron van den and Dieleman, Sander and Schrauwen, Benjamin},
title = {Deep content-based music recommendation},
year = {2013},
publisher = {Curran Associates Inc.},
address = {Red Hook, NY, USA},
booktitle = {Proceedings of the 27th International Conference on Neural Information Processing Systems - Volume 2},
pages = {2643–2651},
numpages = {9},
location = {Lake Tahoe, Nevada},
series = {NIPS'13}
}

@article{schedl2018current,
  author    = {Markus Schedl and Hamed Zamani and Ching-Wei Chen and Yashar Deldjoo and Mehdi Elahi},
  title     = {Current Challenges and Visions in Music Recommender Systems Research},
  journal   = {International Journal of Multimedia Information Retrieval},
  year      = {2018},
  volume    = {7},
  number    = {2},
  pages     = {95--116},
  doi       = {10.1007/s13735-018-0154-2},
  url       = {https://doi.org/10.1007/s13735-018-0154-2},
  issn      = {2192-662X}
}

\end{document}